\documentclass[prl,aps,twocolumn,showpacs,amsmath,amssymb,amsfonts,floatfix,superscriptaddress]{revtex4}
\usepackage{graphicx}
\usepackage{dcolumn}
\usepackage{bm}
\usepackage{latexsym}
\usepackage{float}
\usepackage{supertabular}
\usepackage{longtable}

\begin{document}
\title{Cooperation dilemma in finite populations under fluctuating environments}
\author{Michael Assaf}
\affiliation{Racah Institute of Physics, Hebrew University of Jerusalem, Jerusalem 91904, Israel}
\author{Mauro Mobilia}
\affiliation{Department of Applied Mathematics, School of Mathematics, University of Leeds, Leeds LS2 9JT, U.K.}
\author{Elijah Roberts}
\affiliation{Department of Biophysics, Johns Hopkins University, Baltimore, MD 21218, U.S.A.}
\begin{abstract}
We present a novel approach allowing the study of rare events like fixation under fluctuating environments, modeled as extrinsic noise, in
 evolutionary processes characterized by the dominance of one species.
Our treatment consists of mapping the system  onto an auxiliary model, exhibiting metastable species
 coexistence,  that can be analyzed semiclassically. This approach enables us to study the interplay between extrinsic and demographic noise on the statistics of interest. We illustrate our theory by considering the paradigmatic prisoner's dilemma game whose evolution is described by the probability that cooperators fixate the population and replace all defectors. We analytically and numerically demonstrate that extrinsic noise may drastically enhance the cooperation fixation probability and even change its functional dependence on the population size.
These results, which generalize earlier works in population genetics, indicate that extrinsic noise may help sustain and promote a
 much higher level of cooperation than  static settings.

\end{abstract}

\maketitle

Understanding the origin of cooperative behavior and how it is influenced
by the population's intrinsic properties and by environmental factors are major scientific puzzles~\cite{Pennisi} that are suitably investigated in the framework of evolutionary game theory (EGT)~\cite{EGT}.
In EGT, successful species with a high reproductive potential (fitness) spread and the optimization of the fitness at an individual level can result in the reduction of the population's overall fitness, a
phenomenon suggestively captured by  the prisoner's dilemma (PD) game~\cite{Pennisi,EGT}.
While in EGT the dynamics is traditionally studied in terms of differential equations, demographic fluctuations -- intrinsic noise (IN) -- are known to affect the evolution
in finite populations. In this case, the dynamics is often described by a Markov chain and characterized by the fixation probability of a given trait (or ``pure strategy"), which is  the probability that the trait invades the entire population~\cite{Kimura}. For the classic PD (with IN only), the  cooperation fixation probability (CFP) vanishes exponentially with the population size, see e.g.~\cite{Antal06}, and defection  prevails leading to a cooperation dilemma.
This prediction, at odds with many experimental observations,
has motivated the investigation of various mechanisms that can  promote and sustain cooperation~\cite{various}.

Besides IN, an important source of fluctuations in such systems is extrinsic noise (EN) mostly due to the inherent environmental fluctuations, and from being coupled to other fluctuating systems. Such EN can be aptly modeled in the form of random fluctuations in one or more interaction parameters. In theoretical population genetics~\cite{Kimura,Jensen69,Jensen72,Karlin74}, ecology~\cite{Leigh,KMS08,DT13} and cellular biology~\cite{AR2013}, it has been shown that the combined effect of IN and EN can significantly affect the lifetime of the long-lived metastable coexistence state the system dwells in prior to escape.
In this work, we go beyond these and other works that focused on systems exhibiting metastability, and present a novel approach that allows us to analyze the combined influence of IN and EN, with arbitrary correlation time, magnitude and statistics, in systems characterized by the dominance of one species instead of metastability. This is done by a suitable mapping onto an auxiliary model possessing a long-lived metastable state and treating the latter semiclassically. We illustrate our approach on the prototypical example of the PD game. We show that EN can drastically enhance the CFP and may even change its functional dependence on the population size. These results may be interpreted as the evolutionary signature of noisy environments on population diversity~\cite{bethedging}.

The paradigm of social dilemma is provided by the classic PD,
whose main features are captured by assuming that
the pairwise interaction between cooperators and
defectors is described in terms of the benefit $b$ and cost $c$
 of cooperation, with $b>c>0$~\cite{EGT}.
Here,  mutual cooperation leads to a payoff $b-c>0$ and mutual defection gives a payoff $0$, while when one player defects and the other cooperates, the former gets a payoff $b$ and the latter gets $-c$.
The quantity $r\equiv c/b$ is the cost-to-benefit ratio~\cite{EGT} and the
dilemma arises from the fact that, while $r<1$ and mutual cooperation enhances the population overall payoff,
 each individual is better off defecting.

We consider a finite and well-mixed population of size $N\gg 1$,
with $n$ cooperators  and $N-n$  defectors. At mean field level ($N\to \infty$),
defection always prevails and the fraction $x\equiv n/N$
of cooperators evolves to extinction, $x=0$, according to
the replicator rate equation
$(d/dt)x\equiv \dot{x}\propto x(1-x)\left[\Pi_\textsf{C}(x)-\Pi_\textsf{D}(x)\right]$.  $\Pi_\textsf{C}=bx-c$ and $\Pi_\textsf{D}=bx$
are the cooperator and defector average payoffs, respectively~\cite{EGT}, and we assume that $b,c={\cal O}(1)$.

When the population size is finite, demographic fluctuations always drive
the system to either the absorbing states $n=0$ or $n=N$, and the stochastic
dynamics is described by the master equation $\dot{P}_n=T^{+}_{n-1}P_{n-1}+T^{-}_{n+1}P_{n+1}-(T^{+}_n+T^{-}_n)P_n$, where $T^{+}_n$ and $T^{-}_n$ are the respective birth and death rates. As often, these are given in terms of the Moran model~\cite{Moran,EGT,weaksel,Antal06}: $T^{+}_n=[f_\textsf{C}(n)/\bar{f}(n)]n(N-n)/N^2 $
and $T^{-}_n=[f_\textsf{D}(n)/\bar{f}(n)] n(N-n)/N^2$, where the cooperators/defectors fitnesses are
\begin{eqnarray}\label{fitness}
f_\textsf{C}(n)=1+s\left[bn/N-c\right]\; \text{and}\;
f_\textsf{D}(n)=1+sbn/N,
\end{eqnarray}
and the population average fitness is $\bar{f}=1+s(b-c)n/N$. In Eqs.~(\ref{fitness}) the term $1$ accounts for a baseline fitness contribution and the selection strength is denoted by $s>0$~\cite{EGT,weaksel,MA10}. While our approach applies to arbitrary selection strength, throughout the paper we focus on the biologically relevant limit of weak selection, $s\ll 1$~\cite{Kimura,weaksel}, which ensures that $f_\textsf{C/D}>0$ in Eqs.~(\ref{fitness}).

Furthermore, it is convenient to work in the regime where $s\ll N^{-1/2}$. In this regime, one can accurately approximate the master equation by a Fokker-Planck equation (FPE)~\cite{MA10,Assaf06} for the probability $P(x,t)$ of having cooperator density $x$ at time $t$~\cite{Kimura,TCH}:
\begin{eqnarray}\label{FPE}
\hspace{-1.3mm}\partial_t P(x,t)\!=\!-\partial_x[A(x)P(x,t)]\!+\!1/(2N)\partial^2_x[B(x)P(x,t)],
\end{eqnarray}
where $A(x)=T^+(x)-T^-(x)\sim O(s)$, giving a relaxation time $\propto s^{-1}$,
$B(x)=T^+(x)+T^-(x)$, and  $ T^{\pm}(x)=T_n^{\pm}$.

An important notion to characterize  evolutionary dynamics is the CFP $\phi_\textsf{C}(x_0)$ -- the probability that cooperation fixates starting from an initial fraction $x_0$ of cooperators. In the absence of EN, $\phi_\textsf{C}(x_0)$ can be calculated exactly~\cite{Gardiner,Antal06}, and one finds in the leading exponential order $\phi_\textsf{C}(x_0)\sim  e^{-Nsc(1-x_0)}$.
Here, we purposely  adopt another route and show how to asymptotically calculate $\phi_\textsf{C}(x_0)$ via an auxiliary problem. For this, we consider the modified model obtained by supplementing the original PD system with  a reflecting boundary at $n_0=Nx_0$ by imposing $T^-_{n=n_0}=0$. Hence, the only absorbing state of the modified model is the state $n=N$. Therefore, as $\dot{x}=A(x)<0$ for any $0<x<1$, a quasi-stationary distribution (QSD) peaked at $x_0$ (for any value of  $x_0$) forms  after an  ${\cal O}(s^{-1})$ relaxation time.
This metastable state, however, slowly decays due to a slow leakage of probability into the absorbing state at $x=1$, with a rate given by the inverse of the cooperation  mean fixation time (MFT).

Employing the metastable ansatz  $P(x,t)\simeq \pi(x)e^{-t/\tau}$ in Eq.~(\ref{FPE}), where $\pi(x)$ is the QSD, the MFT $\tau$ of the auxiliary model can be computed using the semiclassical ansatz, $\pi(x)\sim e^{-NS(x)}$. Here $S(x)$ is called the action function, while $p_x(x)\equiv S'(x)$ is the momentum~\cite{Dykman1994,AM10}. This yields a stationary Hamilton-Jacobi equation, $H(x,p_x)=0$, with Hamiltonian $H(x,p_x)=p_xA(x)+(p_x^2/2)B(x)$. Fixation occurs along the  zero-energy trajectory $p_x(x)=-2A(x)/B(x)$, where $p_x(x)\sim {\cal O}(s)\ll 1$. This gives $S(x)=\int p_x dx=(c/b)\ln(2-cs+2bsx)$, from which the QSD at $x>x_0$, $\pi(x)\sim e^{-N[S(x)-S(x_0)]}$, is found. Since $\tau\sim \pi(1)^{-1}$, we have~\cite{AM10,subleading}
\begin{equation}\label{MFT}
\ln \tau \simeq N[S(1)-S(x_0)]\simeq N s c(1-x_0),
\end{equation}
where this result is valid when $Ns\gg 1$, which ensures a long-lived metastable state~\cite{MA10}. Importantly, we find that for $N^{-1}\ll s\ll N^{-1/2}$  the MFT $\tau$ of the modified problem~(\ref{MFT}) coincides to leading  order with the inverse of the CFP in the original PD model~\cite{Antal06,TCH}.  We now use this finding to study the CFP in the presence of EN.

\begin{figure}
\includegraphics[width=3.0in, height=1.5in,clip=]{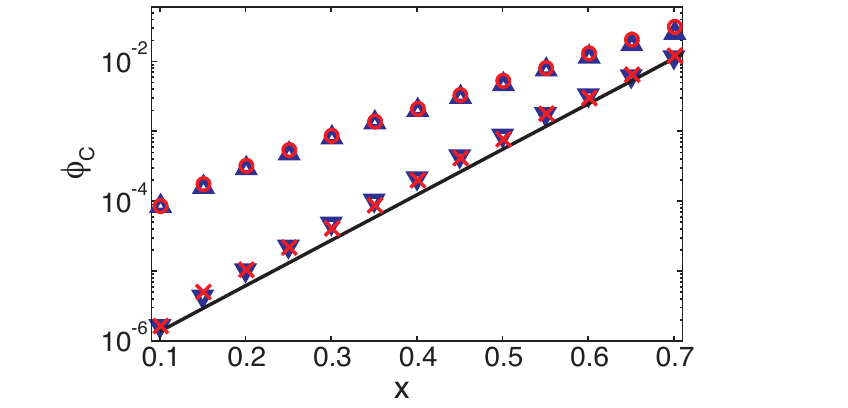}
\caption{(Color online) $\phi_\textsf{C}$ versus $\alpha\tau^{-1}$ for intermediate EN, $\sigma/s_0=0.3$ ($\times$ and $\bigtriangledown$) and strong EN, $\sigma/s_0=1$ ($\bigcirc$ and $\bigtriangleup$). Solid line is the analytical result for $\phi_\textsf{C}$ with IN only. Here $s_0=0.01$, $b=1.25$, $c=1$, $N=1500$, and $\tau_c=20$. The proportionality factor $\alpha$ varies slowly with the model parameters ($35.3$ for intermediate and $55.1$ for strong EN).}
 \label{fig1}
\end{figure}

To this end, we incorporate EN in the form of one or more fluctuating parameters. For concreteness we take a fluctuating selection strength, $s\to s(t)=s_0+\xi(t)$. By directly affecting the fitness of $\textsf{C}/\textsf{D}$ individuals, this choice is particularly relevant in population genetics~\cite{Fisher47,Kimura,Jensen72,Jensen69,Karlin74}, ecology~\cite{ecol} and cellular biology~\cite{bethedging}.
Here, $\xi$ is taken as an Ornstein-Uhlenbeck (OU) process with mean zero, variance $\langle \xi(t) \xi(t') \rangle =\sigma^2 e^{-|t-t'|/\tau_c}$
and correlation time $\tau_c>0$~\cite{Gardiner,OU}. We assume that $\sigma$ is arbitrary so that $s(t)$ can become negative for $\sigma ={\cal O}(s_0)$. The OU process satisfies the following Langevin equation
\begin{eqnarray}\label{lannoise}
\dot{\xi}=-\xi/\tau_c+\sqrt{2\sigma^2/\tau_c}\;\eta(t),
\end{eqnarray}
where $\eta(t)$ is white Gaussian noise
$\langle\!\eta(t) \eta(t')\!\rangle\!=\! \delta(t\!-\!t')$~\cite{eta1}.

We now proceed as in the absence of EN and compute $\tau$ of  the modified  PD model supplemented with a reflecting boundary at $x_0$. We have numerically confirmed (see SM~\cite{SM} for details) that for $Ns_0\gg 1$, $\phi_\textsf{C}(x_0)$ and $\tau^{-1}$ exhibit the same asymptotic behavior in the original and modified models also \textit{in the presence of EN}, see Fig.~\ref{fig1}.

To account for the joint effects of IN and EN, we couple Eq.~(\ref{lannoise}) with FPE~(\ref{FPE}), and arrive at the
 following {\it bivariate} FPE  for the probability  $P(x,\xi,t)$ to find cooperator density $x$ and selection strength  $s=s_0+\xi$ at time $t$:
\begin{eqnarray}\label{2DFPE}
\hspace{-1.3mm}\partial_t P(x,\xi,t)&=&[-\partial_x A+ \partial_{\xi}(\xi/\tau_c)]P(x,\xi,t)\nonumber\\
&+&(2N)^{-1}[\partial^2_x B+(2V/\tau_c) \partial^2_\xi]P(x,\xi,t).
\end{eqnarray}
Here, $A=A(x,\xi)=T^+(x,\xi)-T^-(x,\xi)$ and $B=B(x,\xi)=T^+(x,\xi)+T^-(x,\xi)$ read for $s\ll 1$:
\begin{eqnarray}\label{ABxi}
A(x,\xi)&\simeq&-x(1-x)c(s_0+\xi)[1-(b-c)(s_0+\xi)x],\nonumber\\
B(x,\xi)&\simeq&2x(1-x)[1+c(s_0+\xi)(x-1/2)],
\end{eqnarray}
and we have defined $V\equiv N\sigma^2$. For $N\gg 1$, we can use the semi-classical ansatz for the QSD  $\pi(x,\xi)\sim e^{-NS(x,\xi)}$ in Eq.~(\ref{2DFPE}), which yields the Hamilton-Jacobi equation $H(x,\xi,p_x, p_{\xi})=0$ with Hamiltonian
\begin{equation}\label{ham}
 H=p_x A(x,\xi)-p_{\xi}\xi/\tau_c+(p_x^2/2)B(x,\xi)+(V/\tau_c) p_{\xi}^2,
\end{equation}
where we have defined $p_x=\partial_x S$ and $p_{\xi}=\partial_{\xi} S$.
The corresponding Hamilton equations are
\begin{eqnarray}\label{hameq}
\dot{x}&=&\partial_{p_x} H=A+p_x B\nonumber\\
\dot{p}_x&=&-\partial_x H=-p_x[\partial_x A+(p_x/2)\partial_x B]\nonumber\\
\ddot{\xi}&=& \xi/\tau_c^2 -2(V/\tau_c) p_x \partial_{\xi}A(x,\xi),
\end{eqnarray}
where the third equation has been obtained by combining the equations for $\dot{\xi}$ and $\dot{p}_{\xi}$ into a single
equation  and by keeping terms up to ${\cal O}(p_x)={\cal O}(s_0)$, see below.
 The solution to the Hamilton-Jacobi equation for generic EN (with arbitrary $\tau_c$) is found by solving numerically Eqs.~(\ref{hameq}), yielding the action function $S(x,\xi)=\int p_x(x,\xi)dx+
\int p_{_{\xi}}(x,\xi)d\xi$~\cite{Roma}. Here, we focus on two important and analytically amenable regimes:  {\it short-correlated} (white) EN, when $\tau_c\ll s_0^{-1}$, and  {\it long-correlated} (adiabatic) EN,
when $\tau_c\gg s_0^{-1}$.

For short-correlated EN, $\tau_c\ll s_0^{-1}$, we find that $\ddot{\xi}$ is negligible in the third of Eqs.~(\ref{hameq})~\cite{ddotxi} yielding  the  effective noise strength $\xi \simeq \xi_{\rm eff}\simeq -2cV\tau_c p_x x(1-x)$~\cite{KMS08,Meerson2012,AR2013}. Since $p_x>0$, see below, $\xi_{\rm eff}<0$, thus EN is exploited to enhance the CFP by decreasing the selection strength.

\begin{figure}
\includegraphics[width=3.4in, height=1.7in,clip=]{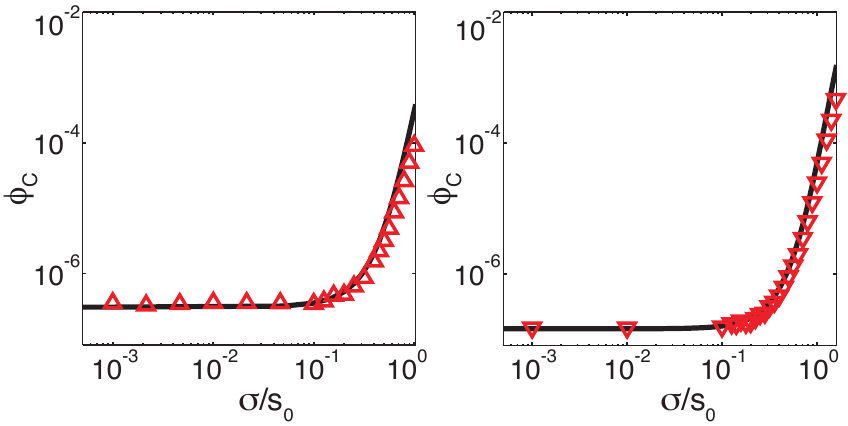}
\caption{(Color online) $\phi_\textsf{C}$ versus relative EN strength $\sigma/s_0$ in the short-correlated EN regime. The solid line is from Eq. (\ref{SandPhi}) and the symbols are numerical simulations. Here, $s_0=0.01$, $b=1.25$, $c=1$ and $N=2000$, $\tau_c=25$, $x_0=0.25$ in the left panel, while $N=1750$, $\tau_c=20$, $x_0=0.1$ in the right panel. The agreement slightly improves from the left to right panels as the inequalities $Ns_0^2\ll 1$ and $\tau_c s_0\ll 1$ are better satisfied.}
 \label{fig2}
\end{figure}

\begin{figure}
\includegraphics[width=2.9in, height=2.1in,clip=]{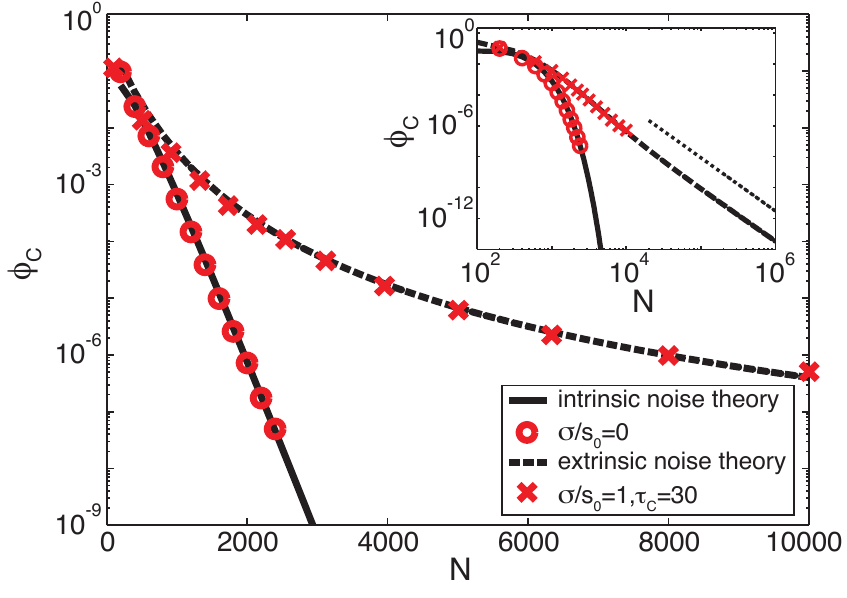}
\caption{(Color online) $\phi_\textsf{C}$ versus $N$ under short-correlated EN: lines are theoretical results while $\times$'s/$\circ$'s are simulation results with/without EN, see legend. Parameters are $s_0=0.01$, $b=1.25$, $c=1$, and $x_0=0.25$. IN-only results display exponential dependence on $N$, whereas for strong EN, $V\tau_c\gg 1$ (see text), $\phi_\textsf{C}$ exhibits  a power-law dependence on $N$. Inset: $\phi_\textsf{C}$ versus $N$ on log-log scale. Results of the theory and simulations are compared to $N^{-3.4}$ (dotted line). The power-law $\phi_\textsf{C}\sim N^{-10/3}$ predicted by~(\ref{SandPhi}) is approached when $N\to\infty$.}
 \label{fig3}
\end{figure}

Substituting $\xi_{\rm eff}$ into the first of Eqs.~(\ref{hameq}) one finds
$\dot{x}= -x(1-x)\left[cs_0-2p_x(1+c^2\tau_c V x(1-x)+{\cal O}(s_0))\right]$.
It appears that EN markedly affects the dynamics when its magnitude satisfies $V \tau_c  \gg {\cal O}(s_0)$. In this regime the corresponding effective white-noise Hamiltonian is $H(x,p_x)\simeq -x(1-x)p_x\left[cs_0-p_x\{1+c^2\tau_c V x(1-x)\}\right]$. Solving $H=0$, we find $p_x=cs_0/(1+c^2\tau_c V x(1-x))$. This yields the MFT in the modified model, and therefore, the CFP of the original PD model:
\begin{eqnarray}
\label{SandPhi}
\hspace{-6mm}
&& \ln \phi_\textsf{C}(x_0)\simeq -N\int_{x_0}^1\,
\frac{cs_0\; du}{1+c^2V \tau_c u(1-u)}= \\
&&-\frac{Ns_0}{c\tau_c V\gamma}\ln{\left\{\left[1+c^2 \tau_c V(1+\gamma)/2\right] \left(\frac{\gamma+1-2x_0}{\gamma-1+2x_0}\right) \right\}}, \nonumber
\end{eqnarray}
where $\gamma=\sqrt{1+4/(c^2\tau_c V)}$. In Fig.~\ref{fig2} we compare Eq.~(\ref{SandPhi}) with numerical simulations as a function of the relative EN strength $\sigma/s_0$ and find a very good agreement for both $x_0= {\cal O}(1)$ (left panel) and $x_0\ll 1$ (right panel). One can clearly see that EN, by effectively decreasing the selection strength $s$, enhances the CFP compared to the IN-only case with $\sigma=0$ (see also Fig.~\ref{fig3} and Fig.~S2 where we respectively plot the CFP versus $N$ and $\tau_c$).

For a given short-correlated EN, $\tau_c\ll s_0^{-1}$, there are two interesting limits to~(\ref{SandPhi}): (i) strong and (ii) weak EN. (i) The most striking effect of EN appears in the limit of strong EN, $V\tau_c\gg 1$, which yields $\gamma \to 1$. Here, for finite values of $x_0>0$, the dependence of $\phi_\textsf{C}(x_0)$ on $N$ becomes a power-law, and Eq.~(\ref{SandPhi}) gives way to $\phi_\textsf{C}(x_0)\!\sim\! \left[ N(\sigma  c)^2 \tau_c (1-x_0)/x_0 \right]^{-(s_0/\sigma^2)/(c\tau_c)}$. This result is confirmed by numerical simulations, see Fig.~\ref{fig3}.
\\(ii) For weak EN, $V\tau_c\ll 1$,  Eq.~(\ref{SandPhi}) can be approximated as $\ln\phi_\textsf{C}(x_0)\simeq -Ns_0 c (1-x_0)\left[1-(1/6)c^2 V\tau_c(1-x_0)(2x_0+1)\right]$, which coincides with the IN-only result to leading order.

The behavior of Eq.~(\ref{SandPhi}) for a small initial density of $\textsf{C}$'s
 ($x_0\ll 1$) is  particularly relevant in  EGT~\cite{EGT}. In this case,
for arbitrary EN strength and $x_0\to 0$, the CFP is
\begin{eqnarray}
\label{phief1}
\ln\phi_\textsf{C}^{(0)}\simeq -[2N s_0/(c\tau_c V\gamma)]\ln \left\{1+c^2\tau_c V(1+\gamma)/2\right\}\!.
\end{eqnarray}
Again, for strong EN, $V\tau_c \gg 1$, Eq.~(\ref{phief1}) becomes a power-law $\phi_\textsf{C}^{(0)}\simeq \left[N(\sigma  c)^2 \tau_c \right]^{-2(s_0/\sigma^2)/(c\tau_c)}$~\cite{other-form}.

Note that, while Eq.~(\ref{SandPhi}) has been formally derived in the regime
$s\ll N^{-1/2}$, its predictions also hold
when $Ns_0^2= {\cal O}(1)$ with $s_0\ll 1$, as illustrated by the numerical results
in Fig.~\ref{fig3}. This is because the leading correction to $\phi_\textsf{C}(x_0)$ due to EN
is independent of $s_0$ when $V\tau_c\gg {\cal O}(s_0)$ [see the denominator of the integrand of
(\ref{SandPhi})]. Thus, our results due to EN are applicable as long as $s_0\ll 1$, and
are expected to hold also in the non-FPE regime where $Ns_0^2\gtrsim {\cal O}(1)$~\cite{SM}.

The case of long-correlated EN, $\tau_c \gg s_0^{-1}$, is investigated in the SM~\cite{SM}.
Here, for weak EN, $V<s_0$, we find that $\ln \phi_\textsf{C}(x_0)\simeq -Ncs_0(1-x_0)\left[1-(c/s_0) V(1-x_0)\right]$.
 Under strong EN, $V>s_0$, the intrinsic fluctuations are negligible~\cite{AR2013} and $\phi_\textsf{C}(x_0)$ is solely
 governed by Eq.~(\ref{lannoise}), yielding
 $\phi_\textsf{C}(x_0)\sim \tau_c^{-1}$, see SM~\cite{SM} for the details.

The various EN parameter regimes for fluctuating $s(t)$ are summarized in a diagram, see Fig.~S1 in the SM~\cite{SM}.

For completeness, we have also considered the case of external fluctuations in the cost-to-benefit ratio  $r=c/b$, with $r\to r(t)=r_0+\xi(t)$ and $r_0<1$ [where $r_0\sim {\cal O}(1)$]. In this case, the dynamics of $\xi$ is given by (\ref{lannoise}) with $\langle \xi(t) \xi(t') \rangle =\sigma_r^2 e^{-|t-t'|/\tau_c}$, where $V_r\equiv N\sigma_r^2$. In addition we assume $\sigma_r\ll r_0$ to guarantee $0<r(t)<1$, and that  $b$ is fixed so that $c(t)=br(t)$ fluctuates.
Performing the calculations along the same lines as for fluctuating $s(t)$, we find for short-correlated EN, $\tau_c\ll s^{-1}$
\begin{equation}\label{phir0}
\ln \phi_\textsf{C}(x_0)\simeq -N\int_{x_0}^{1} \frac{sbr_0du}{1+(sb)^2 V_r \tau_c u(1-u)}.
\end{equation}
Similarly as before, for strong EN $s^2 V_r\tau_c\gg 1$, Eq.~(\ref{phir0}) also predicts that $\phi_\textsf{C}(x_0)$ decays algebraically with $N$.

Our approach generalizes earlier works in population genetics where the combined role of IN and  EN was investigated by considering a fluctuating selection strength, see e.g.~\cite{Fisher47,Jensen69,Jensen72,Karlin74,Kimura}.
In these studies the dynamics was implemented with the Wright-Fisher model with discrete time and non-overlapping
 generations~\cite{Kimura}. In such a setting, a diffusion theory was devised
 in the weak selection limit to account for IN and {\it time-uncorrelated} (white) EN
 by averaging separately on the two sources of noise~\cite{Jensen69,Jensen72,Karlin74,Kimura}. When $N\sigma^2\lesssim N s_0^2\ll 1$, the results of this approach coincide with Eq.~(\ref{SandPhi}) for
 $\tau_c=1$  and $N\to N/2$~\cite{WFM}. Yet, our approach  is more general, since it allows to study EN with arbitrary correlation time and statistics, as well as in the presence of frequency-dependent selection.

In this work, we have analyzed fixation in evolutionary processes  characterized by the dominance of one species.
Our approach relies on a semi-classical treatment applied to an auxiliary model exhibiting metastability.
This allows to study how fixation is affected by the interplay between intrinsic and extrinsic noise (EN).
Our theory is general in the sense that it can deal with EN of arbitrary statistics, correlation time and magnitude, with one or multiple fluctuating parameters, and can be also used for systems exhibiting metastable coexistence.
Using the prototypical prisoner's dilemma game we have shown that EN is exploited to effectively reduce the selection strength and thereby, to drastically enhance  cooperation, whose fixation probability is otherwise vanishingly small. This indicates that EN may be vital in sustaining
a certain level of cooperation and population diversity by effectively opposing single-type dominance, as reported in recent microbial experiments~\cite{bethedging}. Therefore, EN may contribute to reconcile the theoretical predictions with observed examples
of cooperative behaviors.
%
%
%
\renewcommand{\thefootnote}{\fnsymbol{footnote}}
\renewcommand{\thefigure}{S\arabic{figure}}
\renewcommand{\theequation}{S\arabic{equation}}
\setcounter{figure}{0}
\setcounter{equation}{0}


\newpage
\begin{center}
{{\itshape\large Supplemental Material for:}\\~\\\large Cooperation dilemma in finite populations under fluctuating environments}\\~\\
\end{center}

In this supplemental material, we summarize in a schematic diagram
the results obtained  in the main text for the cooperation fixation
probability (CFP) and discuss the various extrinsic noise (EN) parameter regimes. We also  outline the derivation of the CFP under adiabatic EN. Finally, we briefly explain our simulation method.

\section{Parameter regime diagram}
\begin{figure}[ht]
\begin{center}
\includegraphics[width=3.0in, height=2.25in,clip=]{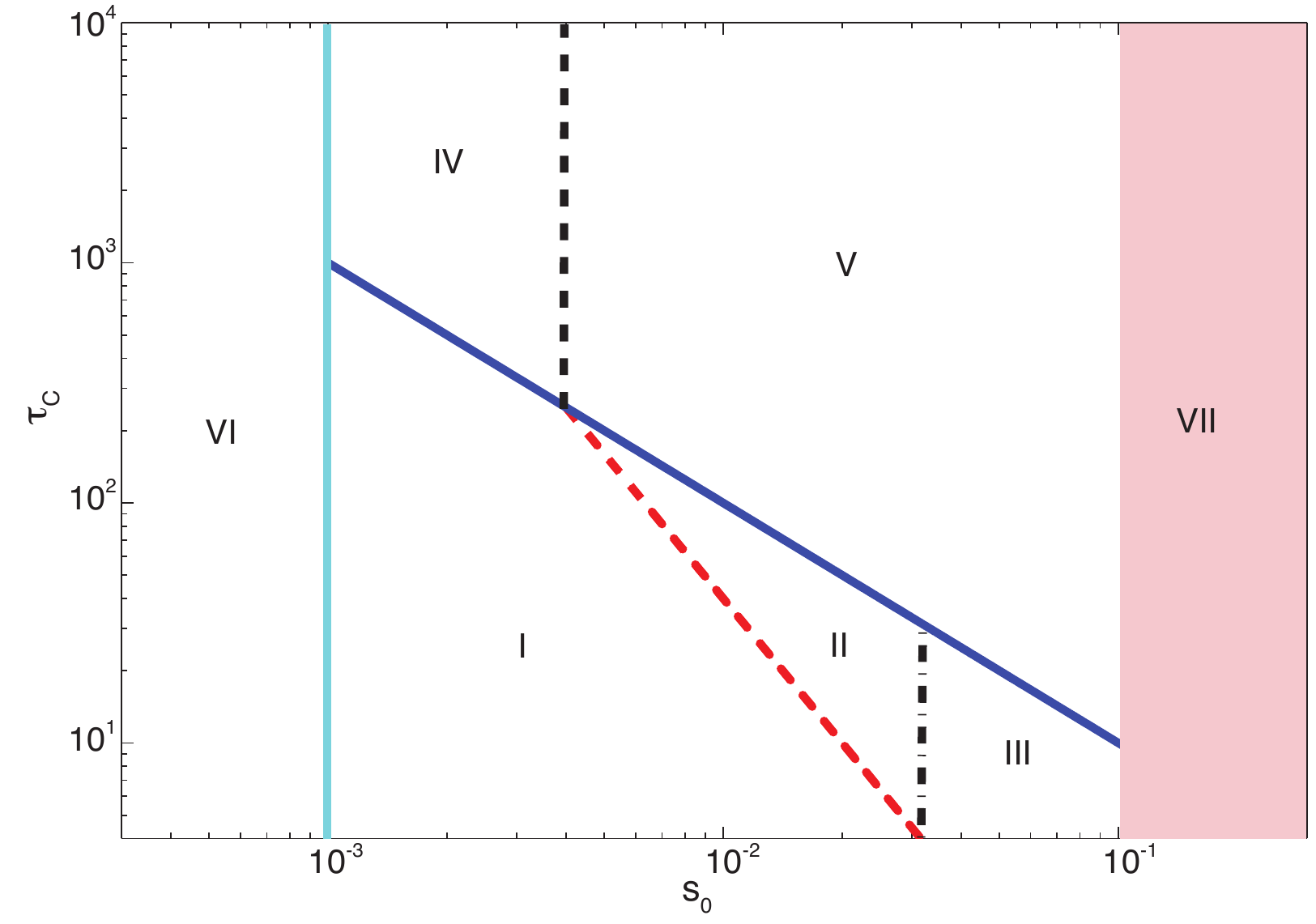}
\end{center}
\caption{{\it (Color online)}. Shown is a schematic diagram $\tau_c$ versus $s_0$ for a fluctuating selection strength $s=s(t)$, when $N=1000$ and $\sigma=0.5s_0$ are kept fixed, so that $V=N\sigma^2=250s_0^2$.
In the main text, analytical results are obtained in the regimes I-V. In these regimes, that are of most physical and biological relevance, EN has been found to greatly enhance the CFP compared to the case with only intrinsic noise (see text, Figs. 2, 3 and \ref{figs2}).
Regimes I-III correspond to short-correlated EN, and are separated by the thick diagonal solid line $\tau_c=1/s_0$ from regimes IV and V, which correspond to long-correlated noise.
Regimes I  and II (weak/strong EN) are separated by the line $\tau_c=1/V$, while Regimes II and III are separated
by a dotted line at $s_0=N^{-1/2}$ and Regimes IV and  V (weak/strong EN) are separated by the line $V=s_0$ (see text). Finally, Regime VI (whose onset is denoted by the thick vertical solid line $s_0=1/N$) and regime VII (the shaded region) are respectively characterized by quasi-neutrality  and strong selection, see the discussion below.}
 \label{figs1}
\end{figure}

In this section we map the results for the CFP in the various regimes of parameter space when the selection strength fluctuates, $s=s(t)=s_0+\xi(t)$. Here, $0<  s_0\leq 1$~\cite{EGT,weaksel} is the mean selection intensity and $\xi(t)$ is the Ornstein-Uhlenbeck (OU) process [Eq.~(4)] with mean zero, variance $\sigma^2$ and correlation time $\tau_c>0$.
When the selection strength fluctuates, in addition to the benefit $b$ and cost $c$ of cooperation, there are four essential parameters that control the system's dynamics: the population size $N$, $s_0$, and the EN magnitude $\sigma$ and correlation time $\tau_c$. In order to present the results in a two-dimensional diagram, we fix $N=1000$ and the relative EN magnitude $\sigma/s_0=0.5$, so that $V\equiv N\sigma^2=250s_0^2$, and draw the schematic diagram of $\tau_c$ versus $s_0$. As shown in Fig.~\ref{figs1}, this diagram is characterized by $7$ distinct regimes which are discussed below.
\\\\

\vspace{0.5cm}

\section{Discussion of the various EN regimes}

{\bf Weak short-correlated noise regime (I)}: Here, $\tau_c\ll s_0^{-1}$, $N^{-1}\ll s_0\ll N^{-1/2}$ and $V\ll 1/\tau_c$. In this regime the CFP $\phi_\textsf{C}$ is given by Eq.~(9), and the dependence on $N$ is exponential.
\\\\
 {\bf Strong short-correlated noise regime (II)}: Here, $\tau_c\ll s_0^{-1}$, $N^{-1}\ll s_0\ll N^{-1/2}$ and $V\gg 1/\tau_c$. In this regime $\phi_\textsf{C}$ is also given by Eq.~(9), but the dependence on $N$ becomes algebraic, see Fig.~3.
\\\\
 {\bf Short-correlated EN / non-Fokker-Planck regime (III)}: Here, $\tau_c\ll s_0^{-1}$ and $N^{-1/2}\lesssim s_0\ll 1$. In this regime  the Fokker-Planck equation is generally not an accurate approximation of the underlying master equation. However, as argued in the main text, and as corroborated by the numerical simulations of Fig.~3 (see the large-$N$ results), the correction to the CFP due to EN predicted by our theory [given by Eq.~(9)] also holds well into the non-Fokker-Planck regime of $Ns_0^2 \gtrsim {\cal O}(1)$.
\\\\
{\bf Weak long-correlated noise regime (IV)}: Here, $s_0^{-1}\ll \tau_c\ll {\cal O}(N/s_0)$ and $V<s_0$. In this regime $\phi_\textsf{C}$ is given by Eq.~(\ref{adiab}), and the dependence on $N$ is exponential as  explained in Section 3 of this supplemental material.
\\\\
{\bf Strong long-correlated noise regime (V)}: Here,  $s_0^{-1}\ll \tau_c\ll {\cal O}(N/s_0)$ and $V>s_0$. In this regime the dynamics is solely governed by the OU process (4) and the CFP satisfies $\phi_\textsf{C}\sim \tau_c^{-1}$ to leading order as explained in Section 3 of this supplemental material (see also Fig.~\ref{figs2} and the main text).
\\\\
 {\bf Quasi-neutral regime (VI)}: Here, $s_0\ll N^{-1}$~\cite{Kimura}.  In this regime the dynamics is close to neutral, and the CFP scales as $\phi_\textsf{C} (x_0)\sim x_0 / N$~\cite{EGT,weaksel}
\\\\
 {\bf Strong-selection regime (VII)}: Here, $s\gtrsim {\cal O}(1)$ (for the sake of illustration, this regime is shown as the shaded region of
 $s_0\gg 0.1$ in Fig.~\ref{figs1}). This strong-selection regime, which is of marginal biological relevance~\cite{Kimura}, can be treated by considering other expressions  [than Eqs.~(1)] for the fitnesses~\cite{EGT,weaksel,Antal06}. In addition, one needs to proceed with  a direct analysis of the master equation (instead of the Fokker-Planck equation), as, e.g., in Ref.~\cite{MA10}.

\vspace{0.5cm}

\section{CFP under long-correlated (adiabatic) EN}
In this section we calculate the CFP under fluctuating selection strength $s=s(t)=s_0+\xi(t)$ in the case of long-correlated (adiabatic) EN, $\tau_c \gg s_0^{-1}$. Here, the selection strength  fluctuates  slowly and can be considered as almost constant while a rare fluctuation leads to the fixation of cooperators. As a result, the CFP can be found by integrating over the fixation probability given noise $\xi_0$, $\phi_\textsf{C}(\xi_0)\sim e^{-cN(s_0+\xi_0)(1-x_0)}$, with the noise's Gaussian weight $e^{-\xi_0^2/(2\sigma^2)}$. A saddle-point approximation gives the optimal noise strength $\xi_0=-cV(1-x_0)$ yielding
\begin{equation}
\label{adiab}
\ln \phi_\textsf{C}(x_0)\simeq -Ncs_0(1-x_0)\left[1-(c/s_0) V(1-x_0)\right].
\end{equation}
This result is valid when $|\ln \phi_\textsf{C}| \gg 1$ which requires $V < s_0$, \textit{i.e.} not too strong EN (Regime IV in Fig.~\ref{figs1}). In addition, since in the original PD model the fixation time is ${\cal O}(N/s_0)$~\cite{Antal06}, the adiabatic regime holds provided that $\tau_c \ll \tau={\cal O}(N/s_0)$. Eq.~(\ref{adiab}) shows that $\phi_\textsf{C}$ can be exponentially enhanced by adiabatic EN.

A different scenario arises under  adiabatic noise of strong intensity (Regime V in Fig.~\ref{figs1}): When $V>s_0$ the intrinsic noise is negligible and the CFP is solely governed by the OU process~(4)~\cite{AR2013}. As a result, the mean fixation time (MFT) in the auxiliary model is determined by the mean first passage time (MFPT) it takes $\xi$ to reach the value  $-s_0$ starting from $\xi=0$ at $t=0$. This MFPT, denoted by $T(\xi)$, is governed by the following equation~\cite{Gardiner}
\begin{equation}
(\sigma^2/\tau_c)T''(\xi)-(\xi/\tau_c)T'(\xi)=-1\nonumber,
\end{equation}
where $\xi \in (-\infty,0]$, and we assume absorbing and reflecting boundaries at $\xi=0$ and $\xi=-\infty$, respectively, such that $T(0)=T'(-\infty)=0$. The solution of this equation is given by
\begin{eqnarray}
T(\xi)&=&\tau_c~{\cal F}(z)\;;\nonumber\\
{\cal F}(z)&=&\left(-\frac{\pi}{2} \mathrm{Erfi}(z)-z^2\,_2\!F_2\left[\{1,1\},\left\{\frac{3}{2},2\right\},z^2\right]\right),\nonumber
\end{eqnarray}
where $\mathrm{Erfi}(z)=(2/\sqrt{\pi})\int_0^{z}e^{y^2}dy$, $\;_2\!F_2(\cdots)$ is the generalized hypergeometric function, $z=\xi/(\sqrt{2}\sigma)$, and ${\cal F}(z)>0$ in the regime of interest $z<0$.

\begin{figure}[ht]
\begin{center}
\includegraphics[width=3.1in, height=2.25in,clip=]{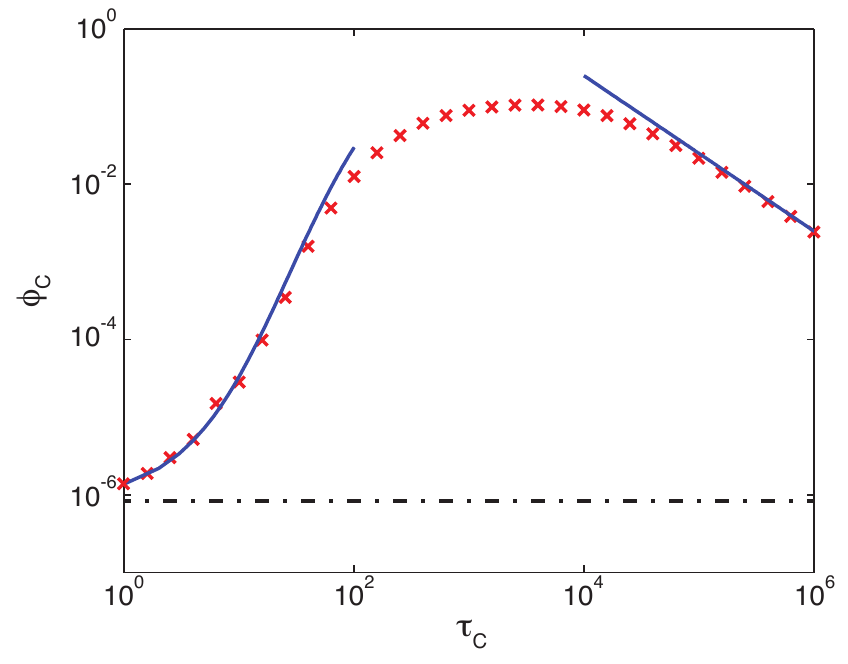}
\end{center}
\caption{{\it (Color online)}. Shown is $\phi_\textsf{C}$ versus $\tau_c$ for $N=2000$, $b=1$, $c=1.25$, $s_0=0.01$, $x_0=0.3$, and strong EN strength with $\sigma/s_0=1$. The left solid line is the theoretical prediction for short-correlated noise [Eq.~(9)], and excellently agrees with simulation results ($\times$'s) up to  $\tau_c={\cal O}(s_0^{-1})$. The right solid line confirms the scaling prediction $\phi_\textsf{C}\sim \tau_c^{-1}$  in the strong adiabatic EN, $\tau_c\gg s_0^{-1}$. The dashed-dotted line corresponds to the CFP in the absence of EN. }
 \label{figs2}
\end{figure}

We are interested in the MFPT to reach noise magnitude $-s_0$ starting from $\xi(0)=0$. Once $\xi$ crosses $-s_0$, the selection pressure vanishes and the auxiliary model rapidly fixates (compared to the fixation time when $s>0$). As a result, the fixation time is governed by $T(\xi=-s_0)$. For strong selection $\sigma={\cal O}(s)$ and for $\xi=-s_0$ we have $z\sim {\cal O}(1)$ and ${\cal F}(z)\sim {\cal O}(1)$. Thus, we find $T(\xi=-s_0)\sim \tau_c$, and as a result, the CFP under strong adiabatic noise in the original problem (see main text) satisfies $\phi_\textsf{C}\sim \tau_c^{-1}$, which is confirmed by Fig.~\ref{figs2}.

\section{Stochastic simulations with EN}
To study fixation we use a kinetic Monte Carlo method based on the next-reaction variant of the Gillespie algorithm~\cite{GillespieGibsonBruck}. During a single trajectory, the current number of cooperators $n$ is stochastically updated using the birth/death transition rates, $NT^{\pm}_n$, described in the main text. EN is added by permitting the selection strength parameter $s$ to fluctuate. A pseudo-reaction fires at intervals much less than the EN correlation time and $s$ is updated as if it had been following an OU process satisfying Eq.~(4) using the method of ~\cite{Gillespie1996ens}. To calculate $\phi_\textsf{C}$, the fraction of many trajectories starting at $n_0$ and resulting in the cooperation state is calculated directly. For the MFT calculation, a reflecting boundary is placed at $n_0$ and the mean time for many ($\geq$1000) trajectories to reach the cooperation state is then calculated.





\begin{thebibliography}{99}
\bibitem{ecol}
Since the work of Fisher and Ford~\cite{Fisher47}, it is conjectured that
noisy selection, $s=s_0+\xi(t)$, can explain the fluctuations other than those produced by
demographic noise~\cite{Kimura,Jensen72,Karlin74}.
Because most fixation properties depend on  the product $Ns_0$ (for $s_0\ll 1$), our findings can also be related to fluctuations in the population size, $N=N(t)$, with $s$ kept fixed, a case particularly relevant for ecology.
\bibitem{OU} Other non-Gaussian statistics are possible. Yet,  without specific knowledge of the EN properties,
we chose the OU noise which is arguably the simplest form of EN with Gaussian statistics and arbitrary correlation time.
\bibitem{eta1}
 $\eta(t)$ is the time-continuous
 limit ($dt \to 0$) of a temporally uncorrelated normal random variable with mean $0$ and
variance $1/dt$~\cite{Gardiner}.
\bibitem{ddotxi} Using the expression of $\xi_{\rm eff}$ together with Eqs.~(\ref{ABxi}) and (\ref{hameq}), one   finds a posteriori that $\ddot{\xi}={\cal O}(V\tau_c s_0^3)\ll  V s_0/\tau_c$  which is negligible  in the third of Eqs.~(\ref{hameq}).
\bibitem{other-form}
It can be checked that Eq.~(\ref{SandPhi}) is well-defined in the double limit of $V\tau_c\gg 1$ ($\gamma \to 1$) and $x_0\to 0$.
\bibitem{WFM}
As time is discrete in the Wright-Fisher model
and the EN between two successive generations is uncorrelated,
the correspondence requires to set $\tau_c=1$ in Eq.~(\ref{SandPhi}). Also, a population of size $2N$ in the Wright-Fisher model maps onto a  population of size $N$ in the Moran model~\cite{Kimura,MA10}.
\end{thebibliography}
\end{document}